\documentclass[12pt,preprint]{aastex}
\usepackage{lineno}
\usepackage{color}
\RequirePackage{lineno}

\def\pcssp{ cm$^{-2}$ s$^{-1}$ }
\def\ltsima{$\; \buildrel < \over \sim \;$}
\def\simlt{\lower.5ex\hbox{\ltsima}} % < over ~
\def\gtsima{$\; \buildrel > \over \sim \;$}
\def\simgt{\lower.5ex\hbox{\gtsima}} % > over ~

\def\deg{\hbox{$^\circ$}}

\def\ltsima{$\; \buildrel < \over \sim \;$}
\def\simlt{\lower.5ex\hbox{\ltsima}} % < over ~
\def\gtsima{$\; \buildrel > \over \sim \;$}
\def\simgt{\lower.5ex\hbox{\gtsima}} % > over ~

\def\deg{\hbox{$^\circ$}}
\def\f{{\em Fermi }}
\def\s{{\em Swift }}
\def\suz{{\em Suzaku }}
\def\XMM{{\em XMM}-Newton }

\def\g{\hbox{$\gamma$}}
\def\l{LAT }

\def\psrbsp{PSR~B1259--63 }

\begin{document}
%\linenumbers 
\title{Discovery of High-Energy Gamma-Ray Emission from the Binary System PSR
  B1259$-$63/LS 2883 Around Periastron with \textit{Fermi}} 
\author{
A.~A.~Abdo\altaffilmark{2,1}, 
M.~Ackermann\altaffilmark{3}, 
M.~Ajello\altaffilmark{3}, 
A.~Allafort\altaffilmark{3}, 
J.~Ballet\altaffilmark{5}, 
G.~Barbiellini\altaffilmark{6,7}, 
D.~Bastieri\altaffilmark{8,9}, 
K.~Bechtol\altaffilmark{3}, 
R.~Bellazzini\altaffilmark{10}, 
B.~Berenji\altaffilmark{3}
% ...
}
\altaffiltext{1}{Corresponding authors: A.~A.~Abdo,
aous.abdo@nrl.navy.mil; S.~Johnston, Simon.Johnston@atnf.csiro.au;
A.~Neronov, Andrii.Neronov@unige.ch; D.~Parent, dmnparent@gmail.com;
K.~Wood, kent.wood@nrl.navy.mil}
\altaffiltext{2}{Center for Earth Observing and Space Research, College of Science, George Mason University, Fairfax, VA 22030, resident at Naval Research Laboratory, Washington, DC 20375}
\altaffiltext{3}{W. W. Hansen Experimental Physics Laboratory, Kavli Institute for Particle Astrophysics and Cosmology, Department of Physics and SLAC National Accelerator Laboratory, Stanford University, Stanford, CA 94305, USA}
% ...
\altaffiltext{5}{Laboratoire AIM, CEA-IRFU/CNRS/Universit\'e Paris Diderot, Service d'Astrophysique, CEA Saclay, 91191 Gif sur Yvette, France}
\altaffiltext{6}{Istituto Nazionale di Fisica Nucleare, Sezione di Trieste, I-34127 Trieste, Italy}
\altaffiltext{7}{Dipartimento di Fisica, Universit\`a di Trieste, I-34127 Trieste, Italy}
\altaffiltext{8}{Istituto Nazionale di Fisica Nucleare, Sezione di Padova, I-35131 Padova, Italy}
\altaffiltext{9}{Dipartimento di Fisica ``G. Galilei", Universit\`a di Padova, I-35131 Padova, Italy}
\altaffiltext{10}{Istituto Nazionale di Fisica Nucleare, Sezione di Pisa, I-56127 Pisa, Italy}
% ...

\begin{abstract}
  We report on the discovery of $\geq$ 100 MeV $\gamma$ rays from the
  binary system PSR B1259$-$63/LS 2883 using the Large Area Telescope
  (LAT) on board \textit{Fermi}. The system comprises a radio pulsar
  in orbit around a Be star. We report on LAT observations from near
  apastron to $\sim60$ days after the time of periastron, $t_p$, on
  2010 December 15. No \g-ray emission was detected from this source
  when it was far from periastron.  Faint $\gamma$-ray emission
  appeared as the pulsar approached periastron.  At $\sim t_p + 30$d, the
  $\geq$ 100 MeV $\gamma$-ray flux increased over a period of a few
  days to a peak flux 20--30 times that seen during the pre-periastron
  period, but with a softer spectrum. For the following month, it was
  seen to be variable on daily time scales, but remained at $\sim 1-4
  \times 10^{-6} \mbox{ cm}^{-2 } \mbox{ s}^{-1 }$ before starting to
  fade at $\sim t_p + 57$d. The total $\gamma$-ray luminosity observed
  during this period is comparable to the spin-down power of the
  pulsar.  Simultaneous radio and X-ray observations of the source
  showed no corresponding dramatic changes in radio and X-ray flux
  between the pre-periastron and post-periastron flares. We discuss
  possible explanations for the observed \g-ray-only flaring of the
  source.
\end{abstract}
 

\section{Introduction}
\label{section-intro}

The pulsar system PSR B1259$-$63 was discovered at Parkes in 1989 and
comprises a 47.76 ms radio pulsar orbiting a massive star (LS 2883) in
a highly elliptical ($e \approx 0.87$) orbit with a period of $\approx
3.4$ years \citep{Johnston1992,Negueruela2011}. Recent optical
spectroscopy \citep{Negueruela2011} yields an updated distance
estimate to this source of $2.3\pm0.4$ kpc, in reasonable agreement
with the dispersion measure (DM) derived distance of 2.7 kpc using the
NE2001 model \citep{Cordes2002}, so we adopt $D=2.3$ kpc.  The
companion shows evidence for an equatorial disk in its optical
spectrum, and has generally been classified as a Be star
\citep{Johnston1994}. The pulsar comes within $\sim 0.67$ AU of its
companion star at periastron, which is roughly the size of the
equatorial disk \citep{Johnston1992}.  The orbital plane of the pulsar
is believed to be highly inclined with respect to this disk and so the
pulsar crosses the disk plane twice each orbit, just before and just
after periastron \citep{melatos1995}.  Shock interaction between the
relativistic pulsar wind and the wind and photon field of the Be star
is believed to give rise to the variable unpulsed X-ray emission
observed throughout the orbit \citep{Cominsky1994, Chernyakova2009}
and the unpulsed radio and TeV $\gamma$ rays observed within a few
months of periastron \citep{ Chernyakova2006}.

At energies around 1 GeV, the Energetic Gamma-Ray Experiment Telescope
({\em EGRET}) provided only an upper limit for the 1994 periastron
passage ($F_{\gamma} \leq 9.4 \times 10^{-8}$ cm$^{-2}$
s$^{-1}$ for E $\geq$ 300 MeV, 95\% confidence,
\citep{Tavani_B1259_1996}).  In TeV \g-rays the system was detected
during the 2004 and 2007 periastron passages and flux variations on
daily timescales were seen for energies $>$ 0.38 TeV in 2004
\citep{Aharonian_B1259_2004_pass, Aharonian_B1259_2007_pass}. 

For the 2010/2011 passage the time of periastron $t_p$ was on 2010
December 15. By comparison to previous passages, the unpulsed radio
and X-ray emission was expected to start rising in mid 2010 November
peaking around $t_p-10$d in the pre-periastron phase and reaching
another peak around $t_p + 15$d in the post-periastron phase. By 2011
April these emissions are expected to go back to their levels when the
pulsar is far from periastron.  

\cite{Atel3085} reported the first discovery of GeV \g-ray emission
from this system which was detected during the first disk passage. A
flaring GeV \g-ray activity during the second disk passage was
reported in \cite{Atel3111} and in \cite{Atel3115}. Recently
\cite{Tam2011} reported with further details the GeV \g-ray activity
from this system. We have assembled a multiwavelength campaign to
monitor the system in radio, optical, X-rays, GeV, and TeV \g-rays
during the 2010/2011 periastron passage. Here we describe the {\em
  Fermi}-LAT detection of PSR B1259$-$63 in the $E\geq 100$ MeV range.
We also present a preliminary analysis of a portion of the radio and
X-ray data to determine if there was any anomalous multiwavelength
behavior compared to previous periastron passages. We have analyzed
LAT data over the entire time period from the beginning of the
\textit{Fermi} mission (2008 August 4; at which time the pulsar was
nearing apastron) through periastron up until 2011 April 22 which is
after the passage of the pulsar through the dense equatorial wind of
the massive star. Full analyses and interpretation of the
multiwavelength data are deferred to subsequent papers.

% In a future paper, we
%will present a more complete analysis of both disk passages with
%comprehensive multiwavelength coverage.  Preliminary results have
%appeared in \cite{Atel3085,Atel3111,Atel3115} and more recently in
%\cite{Tam2011} reported

\section{Observations and Data Analysis}
\label{sec:Observations}

%\psrbsp has a low Galactic latitude, ($l = 304.18^{\circ}, b =
%-0.99^{\circ}$), and therefore careful modeling of Galactic and
%isotropic diffuse \g-ray emission is essential to obtain correct
%fluxes for the sources.  The a
Analysis of the \textit{Fermi} LAT data
was performed using the \textit{Fermi} Science Tools 09-21-00
release. The high-quality ``diffuse'' event class was used together
with the P6$\_$v3$\_$diffuse instrument response functions. To reject
atmospheric \g-rays from the Earth's limb, we selected events with
zenith angle $<100\deg$. We performed standard binned maximum
likelihood analysis using events in the range 0.1--100\,GeV extracted
from a 20$^{\circ} \times 20^{\circ}$ region centered on the location
of PSR~B1259$-$63.  The model includes diffuse emission components as
well as $\gamma$-ray sources within 20$^{\circ}$ of the source (based
on an internal catalog created from 18 months of LAT survey data). The
Galactic diffuse emission was modeled using the
\texttt{gll$\_$iem$\_$v02} model and the isotropic component using
\texttt{isotropic$\_$iem$\_$v02} \footnote{
  http://fermi.gsfc.nasa.gov/ssc/data/analysis/}. To better constrain
the diffuse model components and the nearby sources, we first
generated a model using two years of data between 2008 August 4 and
2010 August 4, a period during which the pulsar was far away from
periastron. We fixed spectral parameters of all the sources between
5$^{\circ}$ and 15$^{\circ}$ from the source, and left free the
normalization factor of all the sources within 5$^{\circ}$ that were
flagged as variable source in the 1FGL catalog
\citep{Fermicatalog}. Normalizations for the diffuse components were
left free as well. For this time period, the source was not detected
with the LAT and we place a 95\% upper limit on the photon flux above
100 MeV F$_{100} < 9 \times 10^{-9} \mbox{ cm}^{-2 } \mbox{ s}^{-1 }$
assuming a power law spectrum with a photon index $\Gamma = 2.1$.

The results of this fit were used to constrain the background source
model for analyses on shorter timescales starting in November 2010. In
the source model, the normalization of the isotropic component was
fixed to the 2-year value, while the normalization for the Galactic
diffuse component and three variable sources were left free.

\begin{figure}[ht]
\begin{center}
\includegraphics[width=0.9\columnwidth]{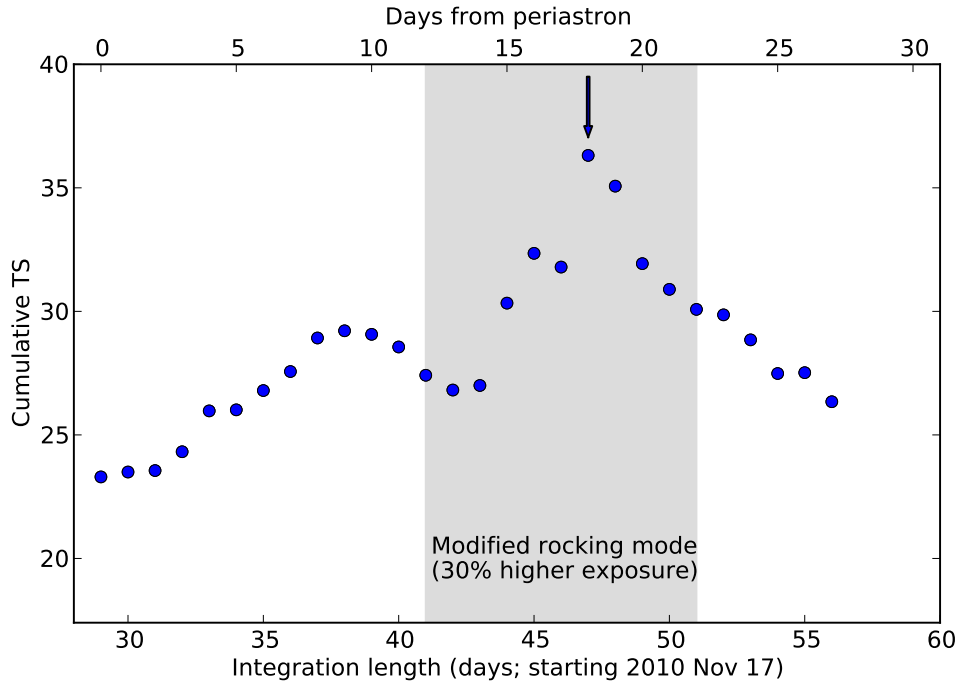}
\caption{Cumulative TS from integrations starting on $t_p-28$d (see text). The shaded
area marks the 10 days during which the LAT was in a modified rocking
mode giving $\sim30\%$ higher exposure on the source. The arrow marks
  the time we adopt as the end of the first emission period.}
\label{fig:TS}
\end{center}
\end{figure}

\begin{figure}[ht]
\begin{center}
\includegraphics[width=0.99\columnwidth]{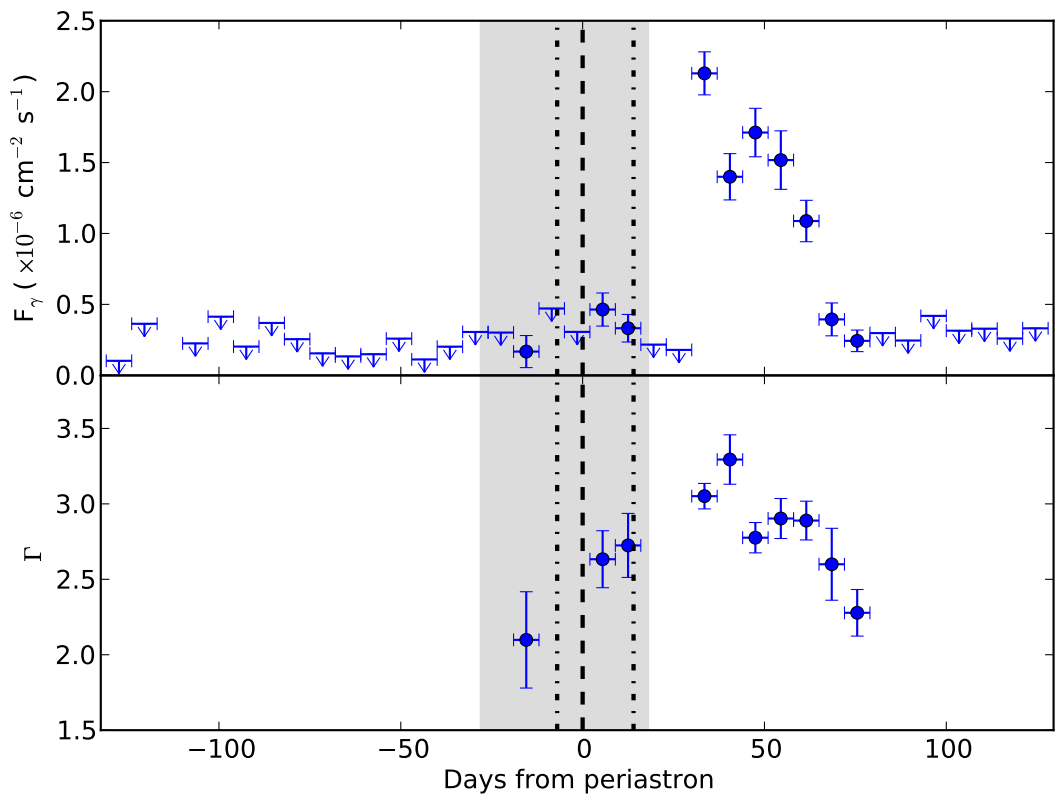}%{B1259_LC_7d_LAT.eps}
\caption{Gamma-ray flux and photon index of \psrbsp in weekly time
  bins between $t-131$d to $t+128$d. \textit{Upper panel}: $\geq$ 100
  MeV flux, 2$\sigma$ upper limits are drawn for points with TS $<$
  5. \textit{Bottom panel}: variations of spectral index of a power
  law spectrum. The shaded area shows the brightening period. Dashed
  line marks the time of periastron. Dashed-dotted lines marks the
  orbital phase during which {\em EGRET} observed this source in 1994
  \citep{Tavani_B1259_1996}.}
\label{fig:LAT_LC}
\end{center}
\end{figure}

We searched for \g-ray emission from this source on daily and weekly
time scales during the first disk passage (mid November to mid
December 2010). No detection at the level of 5$\sigma$ was observed
from the source on these time scales. Integrating from $t_p-28$d (the
typical start of enhanced X-ray and unpulsed radio flux) to periastron
yielded a clear detection of excess \g-ray flux from the source with a
test statistic (TS) of $\sim24$ which corresponds to a detection
significance of $\sim$5$\sigma$ \citep{mattox1996}.  To estimate the
duration of this enhanced emission and to get the best fit for the
spectrum we looked at the cumulative TS as a function of time for
integrations starting at $t_p-28$d (Figure \ref{fig:TS}). Inspection
of this plot reveals that the TS drops monotonically for integrations
ending after $t_p+18$d, so we use this as the end of the integration
for this initial period of detected $\gamma$-ray flux. For the rest of
the paper we will refer to this period as the ``brightening''. \\
During the brightening, the detected \g-ray signal was in the energy
range 0.1--1 GeV and no significant emission was detected above 1 GeV.
The spectrum in the energy range 0.1-1 GeV is best described by a
simple power law with photon index $\Gamma = 2.4 \pm 0.2_\mathrm{stat}
\pm 0.5_\mathrm{sys}$ with average photon and energy
  fluxes of F = $(2.5 \pm 0.8_\mathrm{stat} \pm 0.8_\mathrm{sys})
\times 10^{-7} \mbox{ cm}^{-2 } \mbox{ s}^{-1 }$ and
  F = $( 0.9\pm 0.3_\mathrm{stat} \pm 0.4_\mathrm{sys}) \times
  10^{-10} \mbox{ erg} \mbox{ cm}^{-2 } \mbox{ s}^{-1 }$
  respectively. Because of the low signal-to-noise ratio during this
period spectral fits to an exponentially cutoff power law were not
constraining.  This period is shown as the shaded region on the top
panel of Figure \ref{fig:LAT_LC}, which shows the \g-ray flux in
weekly time bins in the period $t_p-131$d to $t_p+128$d.

At about $t_p+30$d the source brightened rapidly, reaching a flux that
was $\sim10$ times higher than the integrated flux measured during the
first disk passage (Figure \ref{fig:LAT_LC}). This
  flare lasted for about 7 weeks. The spectrum during this
  period ($t_p+30$d to $t_p+79$d) is best described by a power law with
  exponential cutoff. 
%\begin{equation}
%\frac{dN(E)}{dE} = (8.71 \pm 1.17_{stat.} \pm _{sys.}) \times 10^{-10 } E^{-\Gamma} \exp \left(-
%\frac{E}{E_c}\right) \mbox{ cm}^{-2}  \mbox{s}^{-1}  \mbox{MeV}^{-1}  
%\label{eq_plec}
%\end{equation}
  The best-fit result is obtained for a photon index $\Gamma$ = (1.4
  $\pm 0.6_\mathrm{stat} \pm 0.2_\mathrm{sys}$) and a cutoff energy
  E$_c$ = (0.3 $\pm 0.1_\mathrm{stat} \pm 0.1_\mathrm{sys}$) GeV.  The
  average photon and energy fluxes above 100 MeV during this period
  are F$_{100} = $ ($1.3 \pm 0.1_\mathrm{stat} \pm 0.3_\mathrm{sys})
  \times$ 10$^{-6}$ \pcssp and G$_{100} = (4.4 \pm 0.3_\mathrm{stat}
  \pm 0.7_\mathrm{sys}) \times
  10^{-10}  \mbox{ erg} \mbox{ cm}^{-2 } \mbox{ s}^{-1 }$ respectively.  \\
  In addition to the significant difference in flux and spectral shape
  between the brightening and the flare periods, the weekly time bins
  show clear evidence for a change in the photon index for a power-law
  fit on these time scales (bottom panel of Figure \ref{fig:LAT_LC}
  ). The phonon index softens from 2--2.5 during the brightening to a
  value of 3.5 around the peak of the flare. After that, the index
  hardens over the rest of the flare
  period to its values during the brightening.  \\

\begin{figure}[ht]
\begin{center}
\includegraphics[width=0.99\columnwidth]{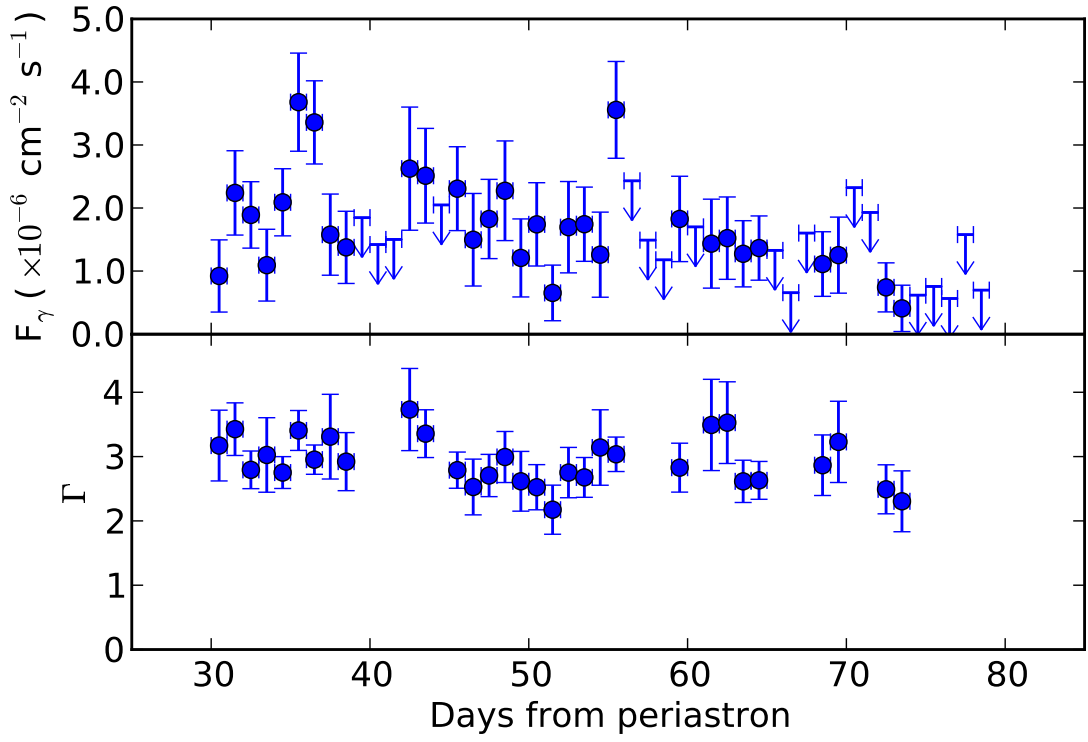}
\caption{Gamma-ray flux and photon index of \psrbsp in daily time bins
  during the flare. \textit{Upper panel}: $\geq$ 100 MeV flux,
  \textit{bottom panel}: spectral index of a power law spectrum.}
\label{fig:LAT_LC_flare}
\end{center}
\end{figure}

The top panel of Figure \ref{fig:LAT_LC_flare} shows \g-ray flux as a
function of time in 1-day time bins during the
flare. The daily source flux during this period is
  variabile at the 99.9\% confidence level according to the method
  outlined in \cite{Fermicatalog}.
\begin{comment}
  To look for flux variability from the source on daily time scales
  during this flare we adopt the method for defining the
  variability index $V$ outlined in \citep{Fermicatalog}. At the
  99.9\% confidence level and for 27 degrees of freedom the light
  curve is inconsistent with being flat if $V > 54.05$. This
  variability condition is met by \psrbsp for which $V=65$ on daily
  time scales during this flaring period.
\end{comment}
During this strong-variability flaring period the source flux
varied by a factor of 2--3 on daily time scales.

The previous upper limits from {\em EGRET} are fully consistent with
the flux observed by \f during the same orbital phase
\citep{Tavani_B1259_1996}. At the time of the bright emission seen by
{\em Fermi}, which was well above the sensitivity level of {\em
  EGRET}, {\em EGRET} was pointed elsewhere.
%in almost the opposite direction from PSR
%B1259$-$63.

\section{Timing Analysis} 

To search for evidence of \g-ray pulsations from PSR B1259$-$63, we
have constructed a radio ephemeris using observations from the 64-m
Parkes telescope.  Using \textsc{Tempo2} \citep{Hobbs2006B}, we fitted
a timing model to 45 TOAs covering the range 2007 Oct 6 through 2011
January 16. Using this ephemeris, we folded LAT photons in the
interval 2008 August 4 through 2010 August 4, a total of 24 months of
observation, all well away from the flaring region near periastron
where non-pulsed \g-ray emission is observed.  We found no
statistically significant indication of a pulsed \g-ray signal from
this source. The pulsed flux upper limit depends on the unknown pulse
shape and assumed spectrum.  We therefore use the continuum upper
limit on the energy flux above 100 MeV of G$_{100} < 1.0 \times
10^{-11}$ erg cm$^{-2}$ s$^{-1}$
%in the
%energy range $\geq$ 100 MeV 
in our comparisons with the rest of the
\g-ray pulsar population.

Comparing this pulsar to the rest of the LAT-detected pulsars, we find
that most detectability metrics predict that this should be a \g-ray
pulsar. Although the characteristic age of 333 kyr is fairly large,
the spin period is short for a middle-aged pulsar and thus at a
distance of 2.3 kpc the $\dot{E}$ of 8.2$\times 10^{35}$ erg s$^{-1}$
and magnetic field at the light cylinder of 2.9 $\times 10^{4}$ G are
well within the range where \g-ray pulsations are typically detected
\citep{LATPSRCAT}. If we assume that the beaming factor $f_\Omega$ is
1, then the \g-ray efficiency of the pulsar is less than 0.7\%, which
is considerably lower than the 10\% efficiencies that are typical in
this $\dot{E}$ range. Determining whether this pulsar is intrinsically
under-luminous in the \g-ray band or if the low \g-ray luminosity is
simply a geometric effect will require detailed modeling that includes
geometrical information from radio polarization measurements.

We have also searched for \g-ray pulsations during the brightening and
flare periods where continuum \g-ray emission was detected. No
pulsations were detected in these intervals. This is consistent with
this \g-ray emission originating from the intrabinary shock, which is well
outside the light cylinder of the pulsar and thus is not expected to
be modulated at the spin period.

\section{Radio and X-ray monitoring}

Pulsed emission was monitored at Parkes to look for changes in the DM
and rotation measure (RM) and determine the duration of eclipse of the pulsed
signal. Pulses disappeared on $t_p-16$d and reappeared on $t_p+15$. In
the $\sim2$ weeks leading up to the disappearance of the pulse,
significant changes in the DM were observed.
 
\begin{figure}[ht]
\begin{center}
\includegraphics[width=0.9\columnwidth]{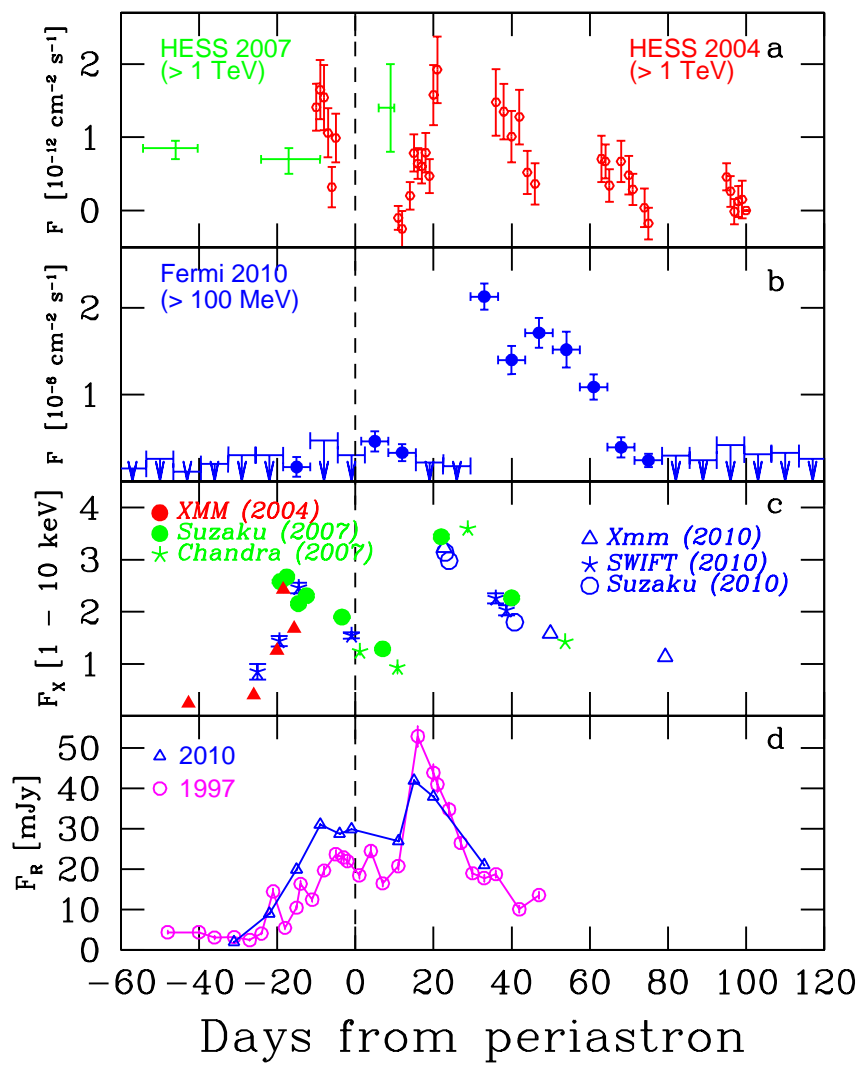}%4lc_4panels_2011.eps}%{xradio_1997.eps}
\caption{Light curves of \psrbsp around periastron. \textit{
    Panel a}: HESS 2004 and 2007 periastron passages
  \citep{Aharonian_B1259_2004_pass}. \textit{Panel b}: \f-\l 2010
  periastron passage. \textit{Panel c}: X-ray fluxes from three
  periastron passages in units of $10^{-11} $ erg cm$^{-2}$ s$^{-1}$
  \citep{Chernyakova2009}. \textit{Panel d}: Radio (2.4 GHz) flux
  densities measured at ATCA for the 2010 and 1997 periastron passages
  \citep{Johnston1999}.}
\label{fig:RADIO-XRAY-FLUX}
\end{center}
\end{figure}

The \psrbsp system was monitored at frequencies between 1.1 and 10 GHz
using the ATCA array. Twelve observations spanning $t_p-31$d  to
$t_p+55$d were collected. Unpulsed transient radio emission was
detected throughout the periastron passage with a behavior similar to
that seen in previous observations \citep{Johnston2005}, as shown in
Figure \ref{fig:RADIO-XRAY-FLUX}.

X-ray observations of \psrbsp during this passage demonstrated the repeatability
of the 1-10 keV light curve as shown in Figure \ref{fig:RADIO-XRAY-FLUX}.
As with the periastron passage
of 2004, \s observed a rapid X-ray brightening starting at
$\sim t_p-25$d. 
These observations confirmed the spectral hardening preceding the
pre-periastron flux rise. 
Similarly to previous
periastron passages (see for example \cite{Chernyakova2009} and
references therein), observations with \s, \suz and \XMM showed a rise
of X-ray flux after periastron.

\section{Discussion} 

Emission from the \psrbsp system is produced in the interaction of the
pulsar wind with the stellar wind of the companion star. Observations
in radio, X-ray and TeV $\gamma$-ray bands
\citep{Johnston1992,Aharonian_B1259_2004_pass,Aharonian_B1259_2007_pass,kawachi04,Tavani_B1259_1996,Chernyakova2006,Chernyakova2009}
revealed a characteristic variability of this emission during the
periods of periastron passage. Detection of the 0.1-10 GeV band
$\gamma$-ray emission around periastron was not unexpected. However,
{\it Fermi} observations reveal puzzling behavior of the source, which
was not predicted in any model of $\gamma$-ray emission from this
system. An unexpected strong flare, visible only in the GeV band was
observed some 30 days after the periastron passage and after the
neutron star passage of the dense equatorial wind of the massive star.

\begin{figure}
\includegraphics[width=0.99\columnwidth]{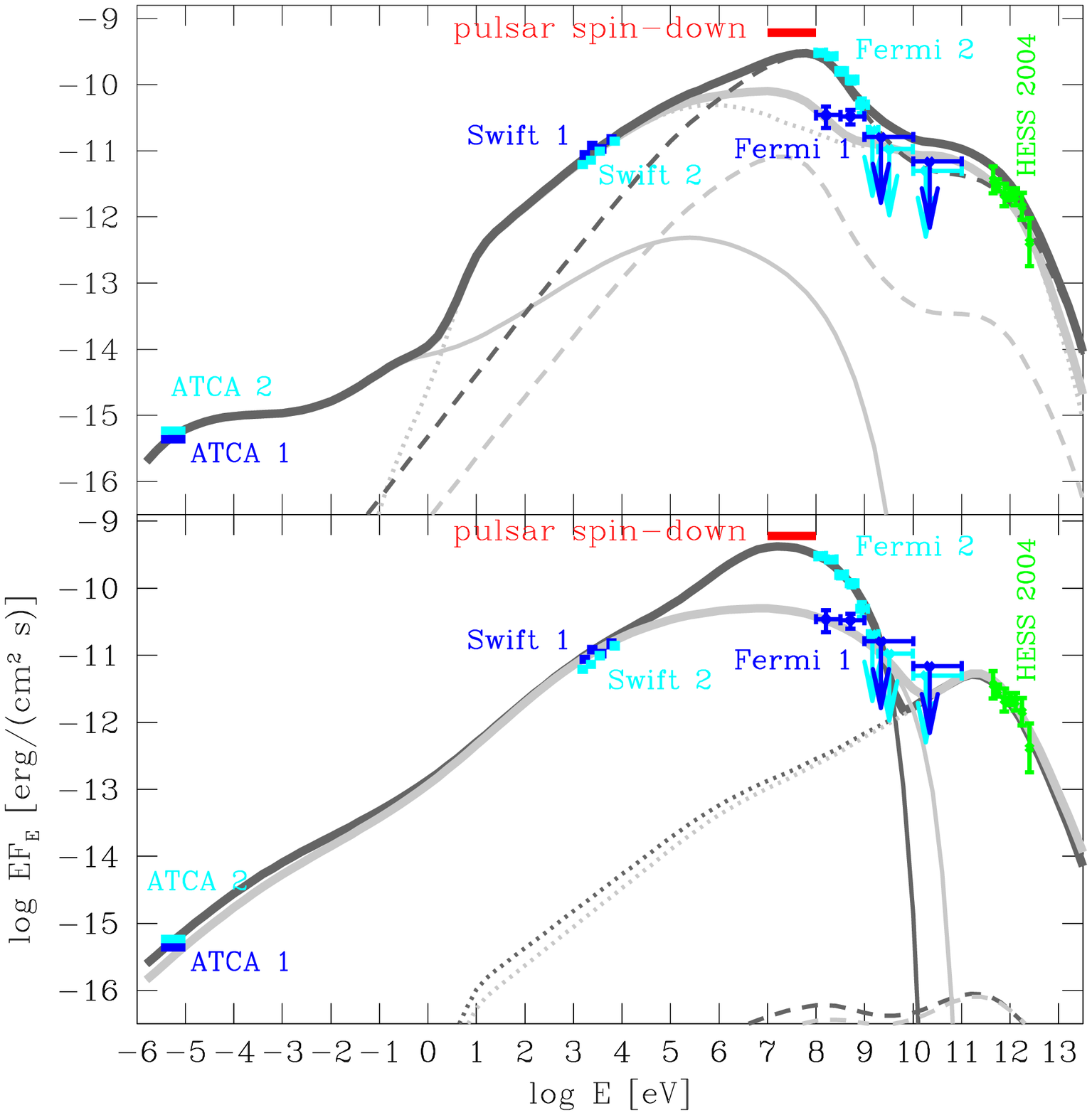}
\caption{Spectral energy distribution of \psrbsp around
  periastron. Blue and cyan points represent the measurements of the
  spectra in the pre- and post-periastron periods (labelled 1 and 2
  respectively) by the LAT, {\em Swift}-XRT in X-rays and ATCA in
  radio. Thin solid, dotted and dashed lines show synchrotron, inverse
  Compton and Bremsstrahlung components correspondingly. Green points
  show HESS measurements from 2004
  \citep{Aharonian_B1259_2004_pass}. Light grey curves show the models
  of pre-periastron emission, dark grey curves show the models of the
  flare. The horizontal red mark shows the flux which would be
  produced if 100\% of the pulsar spin-down power were converted into
  electromagnetic radiation.  In the upper panel the high-energy
  particles are assumed to escape from the system with the speed of
  the stellar wind, while in the lower panel, high-energy particles
  are assumed to escape with the speed $c/3$, as in the model of
  \citet{tavani97}, see text for details. \l data points will be made
  available through https://www-glast.stanford.edu/cgi-bin/pubpub }
\label{fig:synch}
\end{figure}

During this flare the source was characterized by an extremely high
efficiency of conversion of pulsar spin-down power into
$\gamma$-rays. The highest day-average flux was F$_{100} \sim 3.5
\times 10^{-6}$ cm$^{-2 }$ s$^{-1 }$ with a spectral index of
$\Gamma\sim 3.0$. This corresponds to an isotropic $\gamma$-ray
luminosity of $\simeq 8 \times 10^{35}(D/2.3\mbox{ kpc})^2
$~erg~s$^{-1}$, nearly equalling the estimated total pulsar spin down
luminosity $L_{\rm SD}\simeq 8.3\times 10^{35}$~erg s$^{-1}$
\citep{Johnston1992}. This is illustrated in Figure \ref{fig:synch}
where the horizontal red line shows the flux which would be produced
when 100\% of the spin-down power is converted into radiation emitted
within one decade of energy, not taking into account possible beaming
effects.

Broadband spectra of emission around periastron are shown in Figure
\ref{fig:synch}.  Strong increases in GeV flux and changes in \g-ray
spectrum during the flare were not accompanied by noticeable spectral
variations in the X-ray band.

%\textcolor{blue}{ As mentioned above, the possibility of strong
%  post-periastron flare was not considered in any of the existing
%  theoretical models of $\gamma$-ray emission from the source, so that
%  there is no straightforward theoretical interpretation of the
%  observed phenomenon.  Below we discuss possible physical mechanisms
%  which might be responsible for the flare and for the persistent
%  $\gamma$-ray emission during the periastron passage period.}

Several possible mechanisms of production of 0.1-10~GeV $\gamma$-ray
emission from the system were previously discussed: synchrotron,
inverse Compton (IC), Bremsstrahlung, or pion decay emission
\citep{tavani97,kawachi04,Chernyakova2006,khangulyan07}. Electrons
with energies $E_e\sim 100$~TeV produce synchrotron emission in the
energy range $E_\gamma\sim 10^9\left[B/1\mbox{
    G}\right]\left[E_e/10^{14}\mbox{ eV}\right]^2$~eV. Alternatively,
electrons with energies $E_e\sim 1-10$~GeV could produce $\gamma$
quanta with energies $E_\gamma\simeq 10^8\left[E_e/1\mbox{
    GeV}\right]^2$~eV via IC scattering of Be star photons.
Bremsstrahlung emission in the GeV band could be produced by the GeV
electrons. Finally, the dense equatorial stellar wind could provide a
sufficiently dense target for proton-proton interactions followed by
decays of neutral pions into photons.

Figure \ref{fig:synch} shows example model fits to the persistent emission data. The model shown in the upper  panel assumes that 
high-energy particles  escape with the speed of the stellar wind, as in the model of \citet{Chernyakova99} and \citet{Chernyakova2006}. Slow
escape of the high-energy particles leaves enough time for the
efficient cooling of electrons via IC and/or Bremsstrahlung / Coulomb
loss mechanisms.  In the lower panel, high-energy particles are
assumed to escape with the speed $10^{10}$~cm s$^{-1}$, as in the model of
\citet{tavani97}. In this case only synchrotron cooling is efficient.  
The code used for the calculations is described in \citet{zdziarsky10}.

In general, the flare could be explained either by anisotropy of the
$\gamma$-ray emission, or by an abrupt change of physical conditions
in the emission region or by the appearance of a new emission
component.

Several possible sources of anisotropy are present in the system:
relativistic beaming of the $\gamma$-ray emitting outflow from the
system \citep{bogovalov08,dubus10}, anisotropy of pulsar wind, or
anisotropy of radiation field of the massive star.  The model for the
flare spectrum (cyan data points) shown in the lower panel of
Fig. \ref{fig:synch} assumes a particular type of anisotropy which
could appear at the high-energy end of the synchrotron spectrum, at
the energies at which the synchrotron cooling distance is comparable
to the gyroradius. In such a situation the electron distribution could
not be isotropized within the synchrotron cooling time scale. The
assumpiton that highest energy electrons with anisotropic initial
velocity distribution cool before being isotropized, results in the
increase of apparent luminosity by a factor $4\pi/\Omega\sim 10$ where
$\Omega$ is the solid angle into which most of the highest energy
synchrotron power is emitted. 

Another possibility to explain the flare is considered in the model
shown in the upper panel of Fig. \ref{fig:synch}. It iassumes a local
increase of the density of stellar wind by a factor of $\sim 10$ which
results in the increase of the Bremsstrahlung component of emission
spectrum.  
% In the model calculations shown in
%lower panel of Figure \ref{fig:synch}, we assumed a power law injection
%spectrum of electrons up to $2\times 10^{14}$~eV, injected along the
%bow-shock with its apex at the distance $2\times 10^{12}$~cm from the Be
%star, with 1~G magnetic field at the apex.  In the
%calculations of the model shown in the upper panel, an $E^{-2}$
%injection spectrum is assumed.

Clarification of the physical mechanism of the puzzling flare
discovered by {\it Fermi} requires more complete view of the
properties of the flare, including information on system behaviour in
optical and TeV bands and on orbit-to-orbit variations of the GeV
flaring pattern.

%%%%%%%%%%%%%%%%%%%%%%%%%%%%%%%%%%%%%%%%%%%%%%%%%%
%%%%%%%%%%%%%%%%%%%%%%%%%%%%%%%%%%%%%%%%%%%%%%%%%%

\acknowledgments 
We thank Mallory Roberts for helpful contributions.

The $Fermi$ LAT Collaboration acknowledges support
from a number of agencies and institutes for both development and the
operation of the LAT as well as scientific data analysis. These
include NASA and DOE in the United States, CEA/Irfu and IN2P3/CNRS in
France, ASI and INFN in Italy, MEXT, KEK, and JAXA in Japan, and the
K.~A.~Wallenberg Foundation, the Swedish Research Council and the
National Space Board in Sweden. Additional support from INAF in Italy
and CNES in France for science analysis during the operations phase is
also gratefully acknowledged. The Parkes radio telescope is part of
the Australia Telescope which is funded by the Commonwealth Government
for operation as a National Facility managed by CSIRO. We thank our
colleagues for their assistance with the radio timing
observations. This work was supported in part by a NASA \f Guest
Investigator Program.

%\bibliographystyle{apj}
%\bibliography{psrb1259_v9.1}

%\bibliographystyle{apj}
%\bibliography{Pulsar_Catalog_ALL_Refs_new}

\newpage
\begin{thebibliography}{28}
\expandafter\ifx\csname natexlab\endcsname\relax\def\natexlab#1{#1}\fi

\bibitem[{{Abdo} {et~al.}(2010{\natexlab{a}}){Abdo}, {Ackermann}, {Ajello},
  {Allafort}, {Antolini}, {Atwood}, {Axelsson}, {Baldini}, {Ballet},
  {Barbiellini}, \& et~al.}]{Fermicatalog}
{Abdo}, A.~A. {et~al.} 2010{\natexlab{a}}, \apjs, 188, 405

\bibitem[{{Abdo} {et~al.}(2010{\natexlab{b}}){Abdo}, {Ackermann}, {Ajello},
  {Atwood}, {Axelsson}, {Baldini}, {Ballet}, {Barbiellini}, {Baring},
  {Bastieri}, \& et~al.}]{LATPSRCAT}
---. 2010{\natexlab{b}}, \apjs, 187, 460

\bibitem[{{Abdo} {et~al.}(2011){Abdo}, {Parent}, {Dubois}, \&
  {Roberts}}]{Atel3115}
{Abdo}, A.~A., {Parent}, D., {Dubois}, R., \& {Roberts}, M. 2011, The
  Astronomer's Telegram, 3115, 1

\bibitem[{{Abdo} {et~al.}(2010{\natexlab{c}}){Abdo}, {Parent}, {Grove},
  {Caliandro}, {Roberts}, {Johnston}, \& {Chernyakova}}]{Atel3085}
{Abdo}, A.~A., {Parent}, D., {Grove}, J.~E., {Caliandro}, G.~A., {Roberts}, M.,
  {Johnston}, S., \& {Chernyakova}, M. 2010{\natexlab{c}}, The Astronomer's
  Telegram, 3085, 1

\bibitem[{{Aharonian} {et~al.}(2009){Aharonian}, {Akhperjanian}, {Anton},
  {Barres de Almeida}, \& et~al.}]{Aharonian_B1259_2007_pass}
{Aharonian}, F., {Akhperjanian}, A.~G., {Anton}, G., {Barres de Almeida}, U.,
  \& et~al., B. 2009, \aap, 507, 389

\bibitem[{{Aharonian} {et~al.}(2005){Aharonian}, {Akhperjanian}, {Aye},
  {Bazer-Bachi}, \& et~al.}]{Aharonian_B1259_2004_pass}
{Aharonian}, F., {Akhperjanian}, A.~G., {Aye}, K., {Bazer-Bachi}, \& et~al.
  2005, \aap, 442, 1

\bibitem[{{Bogovalov} {et~al.}(2008){Bogovalov}, {Khangulyan}, {Koldoba},
  {Ustyugova}, \& {Aharonian}}]{bogovalov08}
{Bogovalov}, S.~V., {Khangulyan}, D.~V., {Koldoba}, A.~V., {Ustyugova}, G.~V.,
  \& {Aharonian}, F.~A. 2008, \mnras, 387, 63

\bibitem[{{Chernyakova} {et~al.}(2009){Chernyakova}, {Neronov}, {Aharonian},
  {Uchiyama}, \& {Takahashi}}]{Chernyakova2009}
{Chernyakova}, M., {Neronov}, A., {Aharonian}, F., {Uchiyama}, Y., \&
  {Takahashi}, T. 2009, \mnras, 397, 2123

\bibitem[{{Chernyakova} {et~al.}(2006){Chernyakova}, {Neronov}, {Lutovinov},
  {Rodriguez}, \& {Johnston}}]{Chernyakova2006}
{Chernyakova}, M., {Neronov}, A., {Lutovinov}, A., {Rodriguez}, J., \&
  {Johnston}, S. 2006, \mnras, 367, 1201

\bibitem[{{Chernyakova} \& {Illarionov}(1999)}]{Chernyakova99}
{Chernyakova}, M.~A., \& {Illarionov}, A.~F. 1999, \mnras, 304, 359

\bibitem[{{Cominsky} {et~al.}(1994){Cominsky}, {Roberts}, \&
  {Johnston}}]{Cominsky1994}
{Cominsky}, L., {Roberts}, M., \& {Johnston}, S. 1994, \apj, 427, 978

\bibitem[{{Cordes} \& {Lazio}(2002)}]{Cordes2002}
{Cordes}, J.~M., \& {Lazio}, T.~J.~W. 2002, ArXiv e-prints,
  (arXiv:astro-ph/0207156)

\bibitem[{{Dubus} {et~al.}(2010){Dubus}, {Cerutti}, \& {Henri}}]{dubus10}
{Dubus}, G., {Cerutti}, B., \& {Henri}, G. 2010, \aap, 516, A18+

\bibitem[{{Hobbs} {et~al.}(2006){Hobbs}, {Edwards}, \&
  {Manchester}}]{Hobbs2006B}
{Hobbs}, G.~B., {Edwards}, R.~T., \& {Manchester}, R.~N. 2006, \mnras, 369, 655

\bibitem[{{Johnston} {et~al.}(2005){Johnston}, {Ball}, {Wang}, \&
  {Manchester}}]{Johnston2005}
{Johnston}, S., {Ball}, L., {Wang}, N., \& {Manchester}, R.~N. 2005, \mnras,
  358, 1069

\bibitem[{{Johnston} {et~al.}(1992){Johnston}, {Manchester}, {Lyne}, {Bailes},
  {Kaspi}, {Qiao}, \& {D'Amico}}]{Johnston1992}
{Johnston}, S., {Manchester}, R.~N., {Lyne}, A.~G., {Bailes}, M., {Kaspi},
  V.~M., {Qiao}, G., \& {D'Amico}, N. 1992, \apjl, 387, L37

\bibitem[{{Johnston} {et~al.}(1994){Johnston}, {Manchester}, {Lyne},
  {Nicastro}, \& {Spyromilio}}]{Johnston1994}
{Johnston}, S., {Manchester}, R.~N., {Lyne}, A.~G., {Nicastro}, L., \&
  {Spyromilio}, J. 1994, \mnras, 268, 430

\bibitem[{{Johnston} {et~al.}(1999){Johnston}, {Manchester}, {McConnell}, \&
  {Campbell-Wilson}}]{Johnston1999}
{Johnston}, S., {Manchester}, R.~N., {McConnell}, D., \& {Campbell-Wilson}, D.
  1999, \mnras, 302, 277

\bibitem[{{Kawachi} {et~al.}(2004){Kawachi}, {Naito}, {Patterson}, {Edwards},
  {Asahara}, {Bicknell}, {Clay}, {Enomoto}, {Gunji}, {Hara}, {Hara}, {Hattori},
  {Hayashi}, {Hayashi}, {Itoh}, {Kabuki}, {Kajino}, {Katagiri}, {Kifune},
  {Ksenofontov}, {Kubo}, {Kushida}, {Matsubara}, {Mizumoto}, {Mori}, {Moro},
  {Muraishi}, {Muraki}, {Nakase}, {Nishida}, {Nishijima}, {Ohishi}, {Okumura},
  {Protheroe}, {Sakurazawa}, {Swaby}, {Tanimori}, {Tokanai}, {Tsuchiya},
  {Tsunoo}, {Uchida}, {Watanabe}, {Watanabe}, {Yanagita}, {Yoshida}, \&
  {Yoshikoshi}}]{kawachi04}
{Kawachi}, A. {et~al.} 2004, \apj, 607, 949

\bibitem[{{Khangulyan} {et~al.}(2007){Khangulyan}, {Hnatic}, {Aharonian}, \&
  {Bogovalov}}]{khangulyan07}
{Khangulyan}, D., {Hnatic}, S., {Aharonian}, F., \& {Bogovalov}, S. 2007,
  \mnras, 380, 320

\bibitem[{{Kong} {et~al.}(2011){Kong}, {Huang}, {Tam}, \& {Hui}}]{Atel3111}
{Kong}, A.~K.~H., {Huang}, R.~H.~H., {Tam}, P.~H.~T., \& {Hui}, C.~Y. 2011, The
  Astronomer's Telegram, 3111, 1

\bibitem[{{Mattox} {et~al.}(1996){Mattox}, {Bertsch}, {Chiang}, {Dingus},
  {Digel}, {Esposito}, {Fierro}, {Hartman}, {Hunter}, {Kanbach}, {Kniffen},
  {Lin}, {Macomb}, {Mayer-Hasselwander}, {Michelson}, {von Montigny},
  {Mukherjee}, {Nolan}, {Ramanamurthy}, {Schneid}, {Sreekumar}, {Thompson}, \&
  {Willis}}]{mattox1996}
{Mattox}, J.~R. {et~al.} 1996, \apj, 461, 396

\bibitem[{{Melatos} {et~al.}(1995){Melatos}, {Johnston}, \&
  {Melrose}}]{melatos1995}
{Melatos}, A., {Johnston}, S., \& {Melrose}, D.~B. 1995, \mnras, 275, 381

\bibitem[{{Negueruela} {et~al.}(2011){Negueruela}, {Rib{\'o}}, {Herrero},
  {Lorenzo}, {Khangulyan}, \& {Aharonian}}]{Negueruela2011}
{Negueruela}, I., {Rib{\'o}}, M., {Herrero}, A., {Lorenzo}, J., {Khangulyan},
  D., \& {Aharonian}, F.~A. 2011, \apjl, 732, L11+

\bibitem[{{Tam} {et~al.}(2011){Tam}, {Huang}, {Takata}, {Hui}, {Kong}, \&
  {Cheng}}]{Tam2011}
{Tam}, P.~H.~T., {Huang}, R.~H.~H., {Takata}, J., {Hui}, C.~Y., {Kong},
  A.~K.~H., \& {Cheng}, K.~S. 2011, ArXiv e-prints

\bibitem[{{Tavani} \& {Arons}(1997)}]{tavani97}
{Tavani}, M., \& {Arons}, J. 1997, \apj, 477, 439

\bibitem[{{Tavani} {et~al.}(1996){Tavani}, {Grove}, {Purcell}, {Hermsen},
  {Kuiper}, {Kaaret}, {Ford}, {Wilson}, {Finger}, {Harmon}, {Zhang}, {Mattox},
  {Thompson}, \& {Arons}}]{Tavani_B1259_1996}
{Tavani}, M. {et~al.} 1996, \aaps, 120, C221+

\bibitem[{{Zdziarski} {et~al.}(2010){Zdziarski}, {Neronov}, \&
  {Chernyakova}}]{zdziarsky10}
{Zdziarski}, A.~A., {Neronov}, A., \& {Chernyakova}, M. 2010, \mnras, 403, 1873

\end{thebibliography}
%\begin{thebibliography}{27}

%\end{thebibliography}

\end{document}